\documentclass[12pt]{article}
\newcommand{\beq}{\begin{equation}}
\newcommand{\eeq}{\end{equation}}
\newcommand{\bra}{\begin{array}}
\newcommand{\era}{\end{array}}

\newcommand{\al}{\alpha}

\newcommand{\de}{\delta}

\newcommand{\ep}{\epsilon}

\usepackage{graphicx}

\usepackage{latexsym}
\author{Jamila Douari\footnote{jdouari@gmail.com}\\ \\
\small\it LTP, Hay Sa\^ada, Ain Taoujdate, Morocco\rm }
\title{Electrified Fluctuations in D1$\bot$D3 and D1$\bot$D5 Systems}
\begin{document}
\maketitle \vspace*{0.5cm} \maketitle \vspace*{0.5cm} PACS: 11.25.-w; 11.25.Uv; 11.15.Kc; 11.27.qd \vskip1cm
Keywords: Strings, Branes, Dyons, Fluctuations, Boundary Conditions. \vspace*{1.5cm}
\section*{Abstract}
\hspace{.3in}We present the physical phenomenon of the subject discussed in the paper \cite{fluc}. In that paper we dealt with the fluctuations of funnel solutions of intersecting D1 and D3 branes and the electric field $E$ was considered as very high value causing the results to be non-physical. In the present work, the variation interval of $E$ is to be $[0,\frac{1}{\lambda}[$. Then, we extend the study to discuss the overall transverse fluctuations of electrified funnel solutions of D1$\bot$D5 system in the flat background. The boundary conditions are found to be Neumann boundary conditions.
\newpage
\section{Introduction}
\hspace{.3in}D-branes are extended objects, topological defects in string theory on which string endpoints can live. These objects have brought about significant advances in string theory. Various brane configurations have attracted much attention over recent years and several papers have been devoted to the study of the relationship between D-branes with different dimensions \cite{cm,cm2,cm3}. Much of the progress has come about by directly studying the low energy dynamics of the D-branes world volume which is known to be governed by the Born-Infeld (BI) action \cite{BI,InterBran1,InterBran2,dd3}. This world volume theory living on D-branes has many fascinating features. Among these there is the possibility for Dp-branes, through an appropriate excitation of fields, to morph into objects resembling Dp-branes of lower or higher dimensionality. The dynamics of their solutions (bion spike \cite{InterBran1,fun,dual,NBCBI} and fuzzy funnel \cite{fuzfun}) were studied by considering linearized fluctuations around the static solutions \cite{fluct,d1d3}. In the present paper, we go on to study the fluctuations of the fuzzy funnel at the presence of electric field and to limit ourselves to treat the physical phenomenon we consider the electric field in the interval $[0,\frac{1}{\lambda}[$.

The primary goal of this work is to determine the boundary conditions of D1$\bot$D3 and D1$\bot$D5 branes in solving the equations of motion of the fuzzy funnel's fluctuations and discussing the associated potentials. We remark that the variation of the potential $V$ in terms of electric field $E$ and the spatial coordinate $\sigma$ in non-zero modes for both overall and relative transverse fluctuations in D1$\bot$D3 system and in zero mode of overall transverse fluctuations in D1$\bot$D5 system shows a singularity at some stage of $\sigma$. This is more clearly seen at the presence of electric field which leads to separate the system into two regions; small and large $\sigma$ depending on electric field. This implies that these intersecting branes obey Neumann boundary conditions and the end of open string can move freely on the brane. Consequently, the idea that the end of a string ending on a Dp-brane can be seen as an electrically charged particle is supported by the present result. The obtained result in D1$\bot$D3 system is agree with its dual discussed in \cite{NBCBI} considering Born-Infeld action dealing with the fluctuation of the bion spike in D3$\bot$D1 branes case. The dual case of D1$\bot$D5 branes is not yet discussed.

This work is organized as follows: In section 2, we start by a brief review on D1$\bot$D3 and D1$\bot$D5 branes in dyonic case by using the abelian and non-abelian BI actions \cite{NBI,9911136,cm}. In section 3, we study the electrified fluctuations of fuzzy funnel solutions corresponding to D3 and D5 branes. In the first subsection, we review the zero and non-zero modes of the overall transverse electrified fluctuations of fuzzy funnel solutions given in \cite{d1d3}. Then we discuss the zero and non-zero modes of the relative transverse fluctuations in the second subsection. In the third subsection, we treat the electrified fluctuations of the fuzzy funnel in D1$\bot$D5 system. We study the solutions of the linearized equations of motion of the overall transverse fluctuations in zero mode and we discuss its associated potential at the extremities of $\sigma$. The conclusion is presented in section 4.

\section{Intersecting Branes}
\hspace{.3in}In this section, we review in brief the intersection of D1 branes with D3 and D5 branes. We focus our study on the presence of electric field and its influence on the potentials and the fluctuations of the fuzzy funnels.
\subsection{Electrified D1$\bot$D3 System}
\hspace{.3in}We start by giving in brief the known solutions of intersecting D1-D3 branes. From the point of view of D3 brane description the configuration is described by a monopole on its world volume. We use the abelian BI action and one excited transverse scalar in dyonic case to give the bion solution \cite{cm,Gib}. The system is described by the
following action \beq\bra{lll} S=\int dt L &=-T_3 \int
d^4\sigma\sqrt{-det(\eta_{ab}+\lambda^2\partial_a \phi^i \partial_b \phi^i +\lambda F_{ab})}\\\\
&= -T_3 \int d^4\sigma\Big[ 1 +\lambda^2 \Big( \mid \nabla\phi \mid^2 +\stackrel{\rightarrow}{B}^2 +\stackrel{\rightarrow}{E}^2 \Big)\\\\
&+\lambda^4 \Big( (\stackrel{\rightarrow}{B}.\nabla\phi )^2 +(\stackrel{\rightarrow}{E}.\stackrel{\rightarrow}{B})^2 +\mid \stackrel{\rightarrow}{E}\wedge\nabla\phi\mid^2 \Big) \Big]^{\frac{1}{2}} \era\eeq in which
$F_{ab}$ ($a,b=0,...,3$) is the field strength and the electric field is denoted as
$F_{0a}=E_a$. $\sigma^a$ denote the world volume coordinates while $\phi^i$
($i=4,...,9$) are the scalars describing transverse fluctuations of
the brane and $\lambda=2\pi \ell_s^2$ with $\ell_s$ is the string
length. In our case we excite just one scalar so $\phi^i=\phi^9
\equiv\phi$. Following the same process used in the reference
\cite{cm} by considering static gauge, we look for the lowest energy
of the system. Accordingly to (1) the energy of dyonic system is
given as \beq\bra{ll} \Xi&= T_3 \int d^3\sigma\Big[ \lambda^2 \mid
\nabla\phi +\stackrel{\rightarrow}{B}
+\stackrel{\rightarrow}{E}\mid^2 +(1-\lambda^2
\nabla\phi.\stackrel{\rightarrow}{B})^2-2\lambda^2
\stackrel{\rightarrow}{E}.(\stackrel{\rightarrow}{B}
+\nabla\phi)\\\\
&+ \lambda^4 \Big(
(\stackrel{\rightarrow}{E}.\stackrel{\rightarrow}{B})^2 +\mid
\stackrel{\rightarrow}{E} \wedge\nabla\phi\mid^2\Big)
\Big]^{1/2}.\era\eeq Then if we require $\nabla\phi
+\stackrel{\rightarrow}{B} +\stackrel{\rightarrow}{E}=0$, $\Xi$
reduces to \beq\bra{ll} \Xi_0 &=T_3 \int
d^3\sigma\Big[(1-\lambda^2
(\nabla\phi).\stackrel{\rightarrow}{B})^2+2\lambda^2
\stackrel{\rightarrow}{E}.\stackrel{\rightarrow}{E} )\\\\&+
\lambda^4 ( (\stackrel{\rightarrow}{E}.\stackrel{\rightarrow}{B})^2
+{\mid\stackrel{\rightarrow}{E}}\Lambda \nabla\phi\mid^2)
\Big]^{1/2}\era\eeq as minimum energy. By using the Bianchi identity
$ \nabla.\stackrel{\rightarrow}{B}= 0$ and the fact that the gauge field is static, the bion solution is then \beq\phi =\frac{N_m +N_e}{2r},\eeq
with $N_m$ is magnetic charge and $N_e$ electric charge.

Now we consider the dual description of the D1$\bot$D3 branes from D1
branes point of view. To get D3-branes from D-strings, we use the
non-abelian BI action. The natural definition of this action suggested in \cite{STr} is based on replacing the field strength in the BI action by a non-abelian field strength and adding the symmetrized trace $STr(\dots)$ in front of the $\sqrt{det}$ action. The precise prescription proposed in \cite{STr} was that inside the trace one takes a symmetrized average over orderings of the field strength. We refer the reader to \cite{STr} for more details on this action.

The non-abelian BI action describing D-string opening up into a D3-brane is given by
\beq S=-T_1\int d^2\sigma STr \Big[
-det(\eta_{ab}+\lambda^2 \partial_a \phi^i Q_{ij}^{-1}\partial_b
\phi^j )det Q^{ij}\Big]^{1\over2}\eeq where $Q_{ij}=\de_{ij}
+i\lambda \lbrack \phi_i , \phi_j \rbrack$.
Expanding this action to leading order in $\lambda$ yields the usual non-abelian scalar action
$$S\cong -T_1\int d^2\sigma  \Big[ N+ \lambda^2 Tr (\partial_a
\phi^i + \frac{1}{2}\lbrack \phi_i , \phi_j \rbrack \lbrack \phi_j ,
\phi_i \rbrack) +...\Big]^{1\over2}.$$ We deal with the leading order in $N$ when we expand the symmetrized trace and we consider large $N$ limit. The solutions of the equation of motion of the scalar fields $\phi_i$, $i=1,2,3$ represent the
D-string expanding into a D3-brane analogous to the bion solution of
the D3-brane theory \cite{InterBran1,InterBran2}. The solutions are
$$\bra{lc}\phi_i =\pm\frac{\al_i}{2\sigma},&\lbrack \al_i , \al_j
\rbrack=2i\ep^{ijk}\al_k ,\era$$ with the corresponding geometry is
a long funnel where the cross-section at fixed $\sigma$ has the
topology of a fuzzy two-sphere.

The dyonic case is presented by considering ($N, N_f$)-strings. We introduce a background U(1) electric
field on the $N$ D-strings, corresponding to $N_f$ fundamental strings dissolved on the world sheet \cite{dual}. The theory is described by the action \beq S=-T_1\int d^2\sigma STr \Big[
-det(\eta_{ab}+\lambda^2 \partial_a \phi^i Q_{ij}^{-1}\partial_b
\phi^j +\lambda F_{ab})det Q^{ij}\Big]^{1\over2}.\eeq The action can be rewritten as \beq S=-T_1\int d^2\sigma STr \Big[
-det\pmatrix{\eta_{ab}+\lambda F_{ab}& \lambda \partial_a \phi^j
\cr -\lambda \partial_b \phi^i & Q^{ij}\cr}\Big]^{1\over2}.\eeq By computing the determinant, the action becomes \beq
S=-T_1\int d^2\sigma STr \Big[ (1-\lambda^2 E^2 + \al_i \al_i
\hat{R}'^2)(1+4\lambda^2 \al_j \al_j \hat{R}^4 )\Big]^{1\over2},\eeq
in which we replaced the field strength $F_{\tau\sigma}$ by $EI_{N}$ ($I_{N}$ is
$N\times N$-matrix) and the following ansatz were inserted \beq\phi_i =\hat{R}\al_i
.\eeq Hence, we get the funnel solution for dyonic string by solving
the equation of variation of $\hat{R}$ as follows \beq \phi_i
=\frac{\al_i}{2\sigma\sqrt{1-\lambda^2 E^2}}.\eeq

\subsection{Electrified D1$\bot$D5 Branes}
\hspace{.3in}The fuzzy funnel configuration in which the D-strings expand into orthogonal D5-branes shares many common features with the D3-brane funnel. The action describing the static configurations involving five nontrivial scalars is
\beq \bra{ll} S&=-T_1 \int d^2\sigma STr\Big[ 1+\lambda^2
(\partial_\sigma \Phi_i)^2+ 2\lambda^2 \Phi_{ij}\Phi_{ji}+
2\lambda^4 (\Phi_{ij}\Phi_{ji})^2 -4\lambda^4
\Phi_{ij}\Phi_{jk}\Phi_{kl}\Phi_{li}\\\\&\phantom{~~}+2\lambda^4
(\partial_\sigma\Phi_{ij})^2 \Phi_{jk}\Phi_{kj}-4\lambda^4
\partial_\sigma \Phi_i \Phi_{ij}\Phi_{jk}\partial_\sigma \Phi_k
+\frac{\lambda^6 }{4}(\epsilon_{ijklm}\partial_\sigma \Phi_i
\Phi_{jk}\Phi_{lm})^2 \Big]^{1\over 2}, \era\eeq where
$\Phi_{ij}\equiv \frac{1}{2}\lbrack \Phi_i ,\Phi_j \rbrack$ and the
funnel solution is given by suggesting the following ansatz \beq \Phi_i
(\sigma)=\mp\hat{R}(\sigma)G_i,\eeq $i=1,...,5$, where
$\hat{R}(\sigma)$ is the (positive) radial profile and $G_i$ are the
matrices constructed in \cite{funnelSoluD5}. We note that $G_i$ are
given by the totally symmetric $n$-fold tensor product of 4$\times$4
gamma matrices, and that the dimension of the matrices is related to
the integer $n$ by $N=\frac{(n+1)(n+2)(n+3)}{6}$. The Funnel
solution (12) has the following physical radius
\beq R(\sigma)=\frac{\lambda}{N}\sqrt{(Tr(\Phi_i)^2)}=\sqrt{c}\lambda\hat{R}(\sigma),\eeq with $c$ is the "Casimir" associated with the $G_i$ matrices, given by $c=n(n+4)$
and the resulting action for the radial profile $R(\sigma)$ is \beq
S=-NT_1 \int d^2\sigma  \sqrt{1+(R')^2}(1+4\frac{R^4}{c\lambda^2}).\eeq We note that this result only captures the leading large N contribution at each order in the expansion of the square root.

To extend the discussion to dyonic strings we consider ($N,N_f$)-strings. Thus, the electric field is on and the system dyonic is described by the action \beq S=-T_1\int d^2\sigma STr \Big[
-det(\eta_{ab}+\lambda^2 \partial_a \Phi^i Q_{ij}^{-1}\partial_b
\Phi^j +\lambda F_{ab})det Q^{ij}\Big]^{1\over2}.\eeq The action can be rewritten as
$$\phantom{~~~~~~~~~~~~~~~~~~~~~~~}S=-T_1\int d^2\sigma STr \Big[-det\pmatrix{\eta_{ab}+\lambda F_{ab}& \lambda \partial_a \Phi^j \cr -\lambda \partial_b \Phi^i & Q^{ij}\cr}\Big]^{1\over2},\phantom{~~~~~~~~~~~~~~~~~~}(15')$$
with $Q_{ij}=\de_{ij}+i\lambda \lbrack \Phi_i , \Phi_j \rbrack$ and $i,j=1,...,5$, $a,b=\tau,\sigma$. We insert the ansatz (12) and $F_{\tau\sigma}=EI_{N}$ ($I_{N}$ is $N\times N$-matrix) in the action (15). Then we compute the determinant and we obtain
$$\phantom{~~~~~~~~~~~~~~~~~~~~~~~~~~~~~}S=-NT_1 \int d^2\sigma\sqrt{1-\lambda^2 E^2+(R')^2}(1+4\frac{R^4}{c\lambda^2}).\phantom{~~~~~~~~~~~~~~~~~~}(15'')$$
The funnel solution is 
\beq \Phi_i (\sigma)=\mp\frac{R(\sigma)}{\lambda\sqrt{c}}G_i.\eeq
From $(15'')$ We can derive the lowest energy
$$\bra{ll} E&=-NT_1 \int d\sigma\sqrt{\Big(\sqrt{1-\lambda^2 E^2}\pm R'\sqrt{\frac{8R^4}{c\lambda^2}+\frac{16R^8}{c^2\lambda^4}}\Big)^2 +\Big( R'\mp \sqrt{1-\lambda^2 E^2}\sqrt{\frac{8R^4}{c\lambda^2}+\frac{16R^8}{c^2\lambda^4}}\Big)^2}\\
&\ge -NT_1 \int d\sigma\Big(\sqrt{1-\lambda^2 E^2}\pm R'\sqrt{\frac{8R^4}{c\lambda^2}+\frac{16R^8}{c^2\lambda^4}}\Big).\era$$ This is obtained when
$$R'=\mp \sqrt{1-\lambda^2 E^2}\sqrt{\frac{8R^4}{c\lambda^2}+\frac{16R^8}{c^2\lambda^4}}.$$ This equation can be explicitly solved in terms of elliptic functions. For small $R$, the $R^4$ term under the square root dominates, and we find the funnel solution. Then the physical radius of the fuzzy funnel solution (16) is found to be \beq R\approx\frac{\lambda\sqrt{c}}{2\sqrt{2}\sqrt{1-\lambda^2E^2}\sigma}.\eeq

In the next sections, we give an examination of the propagation of the fluctuations on the fuzzy funnel. The setup is similar to both D1$\bot$D5 and D1$\bot$D3 systems. We notice that there are two basic types of funnel's fluctuations, the overall transverse ones in the directions perpendicular to both the Dp-brane (p=3,5) and the string (i.e., $X^{p+1,..,8}$), and the relative transverse ones which are transverse to the string, but parallel to the Dp-brane world volume (i.e., along X$^{1,..,p}$).

\section{Dyonic Funnel's Fluctuations}
\hspace{.3in}In this section, we treat the dynamics of the funnel solutions. We solve the linearized equations of motion for small and
time-dependent fluctuations of the transverse scalars around the exact background in dyonic case.
\subsection{Overall Transverse Fluctuations in D1$\bot$D3 System}
\subsubsection{Zero Mode}
We deal with the fluctuations of the funnel (10) discussed in the previous section. By plugging into the full ($N,N_f$)-string action
(6,7) the "overall transverse" $\delta \phi^m (\sigma,t)=f^m (\sigma,t)I_N$, $m=4,...,8$ which is the simplest type of
fluctuation with $I_N$ the identity matrix, together with the funnel solution, we get \beq\bra{llll} S&=-T_1\int d^2\sigma STr \Big[
(1+\lambda E)(1+\frac{\lambda^2 \al^i \al^i}{4(1-\lambda^2E^2)^2\sigma^4 }) \\&\phantom{~~~~~~~~~}\Big( (1+\frac{\lambda^2 \al^i \al^i}{4(1-\lambda^2E^2)^2\sigma^4})(1-(1-\lambda E)\lambda^2 (\partial_t \delta\phi^m )^2) +\lambda^2 (\partial_\sigma \delta\phi^m )^2\Big) \Big]^{1\over 2}\\\\
&\approx-NT_1\int d^2\sigma H \Big[ (1+\lambda E)-(1-\lambda^2 E^2)\frac{\lambda^2}{2} (\dot{f}^m)^2 +\frac{(1+\lambda E)\lambda^2}{2H} (\partial_\sigma f^m)^2 +...\Big] \era \eeq where $$H=1+\frac{\lambda^2 C}{4(1-\lambda^2E^2)^2\sigma^4}$$ and $C=Tr \al^i \al^i$. For
the irreducible $N\times N$ representation we have $C=N^2 -1$. In the last line we have only kept the terms quadratic in the fluctuations as this is sufficient to determine the linearized equations of motion \beq\Big((1-\lambda E)(1+\lambda^2\frac{N^2-1}{4(1-\lambda^2E^2)^2\sigma^4})\partial^{2}_{t}-\partial^{2}_{\sigma}\Big) f^m =0.\eeq 

In the overall case, all the points of the fuzzy funnel move or fluctuate in the same direction of the dyonic string by an equal distance $\delta x^m$. Thus, the fluctuations $f^m$ could be rewritten as follows \beq f^m (\sigma,t)=\Phi(\sigma)e^{-iwt}\delta x^m,\eeq where $\Phi$ is a function of the spatial coordinate. With this ansatz the equation of motion (19) becomes \beq\Big( (1-\lambda E)(1+\lambda^2\frac{N^2 -1}{4(1-\lambda^2E^2)^2\sigma^4}) w^2 +\partial^{2}_{\sigma}\Big) \Phi(\sigma)=0.\eeq Then, the problem is reduced to finding the solution of a single scalar equation.

In this work, we consider the physical phenomenon which is defined by the fact that the electric field $E$ is in the interval $[0, \frac{1}{\lambda}[$ (contrary to what was treated in \cite{fluc}, such that $E$ was tending to $\infty$).

The equation (21) is an analog one-dimensional Schr\"odinger equation. Let's rewrite it as \beq\Big(\frac{1}{w^2(1-\lambda E)}\partial^{2}_{\sigma}+1+\frac{\lambda^2 N^2 }{4(1-\lambda^2E^2)^2\sigma^4}\Big) \Phi(\sigma)=0,\eeq for large $N$. If we suggest \beq\tilde{\sigma}=w\sqrt{1-\lambda E}\sigma,\eeq the equation (22) becomes \beq\Big(\partial^{2}_{\tilde{\sigma}}+1+\frac{\kappa^2}{\tilde{\sigma}^4}\Big) \Phi(\tilde{\sigma})=0,\eeq with the potential is \beq V(\tilde{\sigma})=\frac{\kappa^2}{\tilde{\sigma}^4}, \eeq and \beq \kappa=\frac{\lambda N w^2}{2(1+\lambda E)}.\eeq The equation (24) is a Schr\"odinger equation for an attractive singular potential $\propto\tilde{\sigma}^{-4}$ and depends on the single coupling parameter $\kappa$ with constant positive Schr\"odinger energy. The solution is then known by making the following coordinate change \beq \chi(\tilde{\sigma})=\int\limits^{\tilde{\sigma}}_{\sqrt{\kappa}} dy\sqrt{1+\frac{\kappa^2}{y^4}}, \eeq and \beq \Phi=(1+\frac{\kappa^2}{\tilde{\sigma}^4})^{-\frac{1}{4}}\tilde{\Phi}. \eeq Thus, the equation (24) becomes \beq\Big( -\partial^{2}_{\chi}+V(\chi)\Big) \tilde{\Phi}=0,\eeq with \beq V(\chi)=\frac{5\kappa^2}{(\tilde{\sigma}^2+\frac{\kappa^2}{\tilde{\sigma}^2})^3}.\eeq

Accordingly to the variation of this potential (Fig.1), the system looks like separated into two regions depending on $\sigma$. In small $\sigma$ region $V$ is close to 0 with a constant value for all $E$. In large $\sigma$ region, specially when $\sigma$ reaches 0.7, $V$ increases too fast as we jump to a new region and gets a maximum value when $E\approx0.5$. 

Then, the fluctuation is found to be \beq \Phi=(1+\frac{\kappa^2}{\tilde{\sigma}^4})^{-\frac{1}{4}}e^{\pm i\chi(\tilde{\sigma})}. \eeq This fluctuation has the following limits; at large $\sigma$, $\Phi\sim e^{\pm i\chi(\tilde{\sigma})}$ and if $\sigma$ is small
$\Phi=\frac{\tilde{\sigma}}{\sqrt{\kappa}}e^{\pm i\chi(\tilde{\sigma})}$. These are the asymptotic wave function in the regions $\chi\rightarrow \pm\infty$, while around $\chi\sim 0$; i.e. $\tilde{\sigma}\sim\sqrt{\kappa}$, $\Phi\sim 2^{-\frac{1}{4}}$. Also we find that $\Phi$ has different expressions in small and large $\sigma$ regions.
\subsubsection{Non-Zero Modes}
The fluctuations discussed above could be called the zero mode $\ell=0$ and for non-zero modes $\ell\geq0$, the fluctuations are $\delta \phi^m (\sigma,t)=\sum\limits^{N-1}_{\ell=0}\psi^{m}_{i_1 ... i_\ell}\al^{i_1} ... \al^{i_\ell} $ with $\psi^{m}_{i_1 ... i_\ell}$ are completely symmetric and traceless in the lower indices.

The action describing this system is \beq\bra{lll}
S&\approx-NT_1\int d^2\sigma  \Big[ (1+\lambda E)H-(1-\lambda^2
E^2)H\frac{\lambda^2}{2} (\partial_{t}\delta\phi^m)^2) \\\\
&+\frac{(1+\lambda E)\lambda^2}{2H} (\partial_\sigma \delta\phi^m)^2
-(1-\lambda^2 E^2)\frac{\lambda^2}{2}\lbrack \phi^i ,\delta\phi^m
\rbrack^2 \\\\
&-\frac{\lambda^4}{12}\lbrack \partial_{\sigma}\phi^i
,\partial_{t}\delta\phi^m \rbrack^2+...\Big] \era\eeq Now the
linearized equations of motion are \beq \Big[(1+\lambda
E)H\partial_{t}^2 -\partial_{\sigma}^2\Big]\delta\phi^m
+(1-\lambda^2 E^2)\lbrack \phi^i ,\lbrack \phi^i ,\delta\phi^m
\rbrack\rbrack -\frac{\lambda^2}{6}\lbrack \partial_{\sigma}\phi^i
,\lbrack \partial_{\sigma}\phi^i ,\partial^2 _{t}\delta\phi^m
\rbrack\rbrack=0,\eeq with $H=1+\lambda^2\frac{N^2 -1}{4(1-\lambda^2E^2)^2\sigma^4}$. Since the background solution is $\phi^i
\propto \al^i$ and we have $\lbrack \al^i , \al^j \rbrack
=2i\epsilon_{ijk}\al^k $, we get \beq\bra{ll} \lbrack \al^i ,
\lbrack\al^i, \delta\phi^m \rbrack
&=\sum\limits_{\ell<N}\psi^{m}_{i_1 ... i_\ell}\lbrack \al^i ,
\lbrack\al^i ,\al^{i_1} ... \al^{i_\ell} \rbrack\\\\
&=\sum\limits_{\ell<N}4\ell(\ell+1)\psi^{m}_{i_1 ...
i_\ell}\al^{i_1} ... \al^{i_\ell} \era\eeq To obtain a specific
spherical harmonic on 2-sphere, we have \beq\lbrack \phi^i ,\lbrack
\phi^i ,\delta\phi_{\ell}^m
\rbrack\rbrack=\frac{\ell(\ell+1)}{(1-\lambda^2E^2)\sigma^2}\delta\phi_{\ell}^m
,\phantom{~~~~~~}\lbrack \partial_{\sigma}\phi^i ,\lbrack
\partial_{\sigma}\phi^i ,\partial_{t}^2 \delta\phi^m
\rbrack\rbrack=\frac{\ell(\ell+1)}{(1-\lambda^2E^2)^2\sigma^4}\partial_{t}^2\delta\phi
_{\ell}^m .\eeq Then for each mode the equations of motion are \beq
\Big[ \Big( (1+\lambda E)(1+\lambda^2\frac{N^2 -1}{4(1-\lambda^2E^2)^2\sigma^4})
-\frac{\lambda^2\ell(\ell+1)}{6(1-\lambda^2E^2)^2\sigma^4}\Big) \partial_{t}^2
-\partial_{\sigma}^2 +\frac{\ell(\ell+1)}{\sigma^2}
\Big]\delta\phi_{\ell}^m =0.\eeq The solution of the equation of
motion can be found by taking the following proposal. Let's consider
$\phi_{\ell}^m =f^m_\ell (\sigma)e^{-iwt}\delta x^m$ in direction
$m$ with $f^m_\ell (\sigma)$ is some function of $\sigma$ for each
mode $\ell$.

The last equation can be rewritten as \beq \Big[-\partial_{\sigma}^2
+V(\sigma) \Big] f_{\ell}^m (\sigma)=w^2 (1+\lambda E) f_{\ell}^m
(\sigma),\eeq with
$$V(\sigma)=-w^2 \Big( (1+\lambda E)\frac{\lambda^2 N^2}{4(1-\lambda^2E^2)^2\sigma^4}
-\frac{\lambda^2\ell(\ell+1)}{6(1-\lambda^2E^2)^2\sigma^4}\Big)+\frac{\ell(\ell+1)}{\sigma^2}.$$

In small $\sigma$ region, this potential is reduced to
$$V(\sigma)= \frac{-w^2\lambda^2}{(1-\lambda^2E^2)^2\sigma^4}\Big( \frac{ (1+\lambda E)N^2}{4}
-\frac{\ell(\ell+1)}{6}\Big).$$
This potential (Fig.2(a)) is close to 0 for almost of $\sigma$ and $E$ until that $E\approx0.87$ we remark that $V$ changes at $\sigma\approx0.04$ and then goes up too fast to be close to 0 again for the other values of $\sigma$.

In small $\sigma$ limit, we reduce the equation (37) to the following form \beq \Big[ w^2 \Big( (1+\lambda E)(1+\lambda^2\frac{N^2 -1}{4(1-\lambda^2E^2)^2\sigma^4})
-\frac{\lambda^2\ell(\ell+1)}{6(1-\lambda^2E^2)^2\sigma^4}\Big)+\partial_{\sigma}^2 \Big]
f_{\ell}^m  (\sigma)= 0.\eeq and again as \beq \Big[ 1+
\frac{1}{(1-\lambda^2E^2)^2\sigma^4}\Big(\lambda^2\frac{N^2
-1}{4}-\frac{\lambda^2\ell(\ell+1)}{6(1+\lambda E)}\Big)+\frac{1}{w^2 (1+\lambda
E)}\partial_{\sigma}^2  \Big] f_{\ell}^m  (\sigma)= 0.\eeq
We define new coordinate $\tilde{\sigma}=w\sqrt{1+\lambda E}\sigma$
and the latter equation becomes \beq
\Big[\partial_{\tilde{\sigma}}^2 + 1+
\frac{\kappa^2}{\tilde{\sigma}^4}\Big] f_{\ell}^m (\sigma)= 0,\eeq where
$$
\kappa^2=\frac{w^2(1+\lambda E)}{(1-\lambda^2E^2)^2}\Big(\lambda^2\frac{N^2
-1}{4}-\frac{\lambda^2\ell(\ell+1)}{6(1+\lambda
E)}\Big)^{\frac{1}{2}}$$
such that $$N>\sqrt{\frac{2\ell(\ell+1)}{3(1+\lambda E)}+1}.$$
By following the same setup of zero mode, we get the solution by using the steps
(27-31) with new $\kappa$. Since we considered small $\sigma$ we get
\beq
V(\chi)=\frac{5\tilde{\sigma}^6}{\kappa^4},
\eeq
and the fluctuation is found to be \beq f^m_\ell=\frac{\tilde{\sigma}}{\sqrt{\kappa}}e^{\pm
i\chi(\tilde{\sigma})}.\eeq in small $\sigma$ region.

Now, let's check the case of large $\sigma$. In this case, the equation of motion (37) of the fluctuation can be rewritten in the following form \beq \Big[-\partial_{\sigma}^2 +V(\sigma) \Big] f_{\ell}^m (\sigma)= w^2 (1+\lambda E) f_{\ell}^m (\sigma),\eeq with
$$V(\sigma)=\frac{\ell(\ell+1)}{\sigma^2}.$$ We remark that, in large $\sigma$ limit (Fig.2(b)), the potential $V$ is independent of $E$ and going down as $\sigma$ is going up. The figures 2(a) and 2(b) show that the system in non-zero modes is separated to two totally different regions and the main remark is that the potential gets a singularity at some level of $\sigma$ which is considered the intersection of small and large $\sigma$ regions. In our calculations we took small $\sigma$ from zero until the half of the unit of $\lambda=1$ and the large $\sigma$ region from 0.5 until 1 with $w=1$, $l=1$ and $N=10$.

The $f_{\ell}^m$ is now a Sturm-Liouville eigenvalue problem. The fluctuation is found to be \beq\bra{ll}f^m_\ell(\sigma)&=\alpha \sqrt{\sigma}BesselJ\Big(\frac{1}{2}\sqrt{1+4\ell(\ell+1)},w\sigma\sqrt{1+\lambda E}\Big)\\ &+\beta\sqrt{\sigma}BesselY\Big(\frac{1}{2}\sqrt{1+4\ell(\ell+1)},w\sigma\sqrt{1+\lambda E}\Big),\era\eeq with $\alpha$, $\beta$ are constants. Again, it's clear that the fluctuation solution in this case is totally different from the one gotten in small $\sigma$ limit (43) supporting the idea that the system is divided to two regions. In the following, we continue the study of D1$\bot$D3 branes by dealing with the relative transverse fluctuations.

\subsection{Relative Transverse Fluctuations in D1$\bot$D3 System}
\subsubsection{Zero Mode}
\hspace{.3in}In this subsection, we consider the "relative transverse" fluctuations $\delta
\phi^i (\sigma,t)=f^i (\sigma,t)I_N$, $i=1,2,3$, and the action describing the system has the expression \beq S=-T_1\int d^2\sigma STr \Big[ -det\pmatrix{\eta_{ab}+\lambda F_{ab}& \lambda \partial_a (\phi^j +\delta\phi^j )\cr -\lambda \partial_b (\phi^i +\delta\phi^i )& Q^{ij}_*
\cr}\Big]^{1\over2},\eeq with $$Q^{ij}_* =Q^{ij}+i\lambda (\lbrack \phi_i,\delta\phi_j \rbrack+\lbrack \delta\phi_i,\phi_j
\rbrack+\lbrack \delta\phi_i,\delta\phi_j \rbrack).$$ As done above, we keep only the terms quadratic in the fluctuations and the action becomes \beq S\approx-NT_1\int d^2\sigma  \Big[ (1-\lambda^2
E^2)H-(1-\lambda E)\frac{\lambda^2}{2}(\dot{f}^i)^2+\frac{(1+\lambda E)\lambda^2}{2H} (\partial_\sigma
f^i)^2 +...\Big],\eeq with $H=(1+\lambda^2\frac{N^2 -1}{4(1-\lambda^2E^2)^2\sigma^4})$.

Then we define the relative transverse fluctuation as $f^i =\Phi^i (\sigma)e^{-iwt}\de x^i$ in the direction of $x^i$, with $\Phi$ is a function of $\sigma$, and the equations of motion of the fluctuations are found to be \beq\Big ( -\partial^{2}_{\sigma}-\frac{w^{2}\lambda^2(1-\lambda E)(N^2-1)}{4(1+\lambda E)(1-\lambda^2E^2)^2\sigma^4}\Big)\Phi^i = w^2 \frac{1-\lambda E}{1+\lambda E}\Phi^i,\eeq where the potential is $$V(\sigma)=-\frac{w^{2}\lambda^2(1-\lambda E)(N^2-1)}{4(1+\lambda E)(1-\lambda^2E^2)^2\sigma^4}.$$
We remark that the presence of E is quickly increasing the potential from $-\infty$ to zero. Then, when $E$ is close to the inverse of $\lambda$ the potential is close to zero for all $\sigma$;
\begin{itemize} \item $E\sim 0$, $V(\sigma)\sim -\lambda^2\frac{N^2
-1}{4\sigma^4}w^{2}$ \item $E\sim \frac{1}{\lambda}$, $V(\sigma)\sim
-\frac{1-\lambda E}{2}\lambda^2\frac{N^2-1}{4(1-\lambda^2E^2)^2\sigma^4}w^{2}$.
\end{itemize}
This case is seen as a zero mode of what is following so we will focus on its general case known as non-zero modes.
\subsubsection{Non-Zero Modes}
\hspace{.3in}Let's give the equation of motion of relative transverse fluctuations of non-zero $\ell$ modes with ($N,N_f$)-strings intersecting D3-branes. The fluctuation is given by $\delta \phi^i
(\sigma,t)=\sum\limits^{N-1}_{\ell=1}\psi^{i}_{i_1 ...i_\ell}\al^{i_1} ... \al^{i_\ell} $ with $\psi^{i}_{i_1 ... i_\ell}$ are completely symmetric and traceless in the lower indices.

The action describing this system is \beq\bra{lll}
S&\approx-NT_1\int d^2\sigma  \Big[ (1-\lambda^2 E^2 )H-(1-\lambda
E)H\frac{\lambda^2}{2} (\partial_{t}\delta\phi^i)^2) \\\\
&+\frac{(1+\lambda E)\lambda^2}{2H} (\partial_\sigma \delta\phi^i)^2
-(1-\lambda E)\frac{\lambda^2}{2}\lbrack \phi^i ,\delta\phi^i
\rbrack^2 \\\\
&-\frac{\lambda^4}{12}\lbrack \partial_{\sigma}\phi^i
,\partial_{t}\delta\phi^i \rbrack^2+...\Big]. \era\eeq The equation
of motion for relative transverse fluctuations in non-zero modes is
\beq \Big[\frac{1-\lambda E}{1+\lambda E}H\partial_{t}^2
-\partial_{\sigma}^2\Big]\delta\phi^i +(1-\lambda E)\lbrack \phi^i
,\lbrack \phi^i ,\delta\phi^i \rbrack\rbrack
-\frac{\lambda^2}{6}\lbrack \partial_{\sigma}\phi^i ,\lbrack
\partial_{\sigma}\phi^i ,\partial^2 _{t}\delta\phi^i
\rbrack\rbrack=0.\eeq
By the same way followed in overall case the
equation of motion for each mode $\ell$ is found to be \beq \Big[
-\partial_{\sigma}^2+\Big( \frac{1-\lambda E}{1+\lambda
E}(1+\lambda^2\frac{N^2 -1}{4(1-\lambda^2E^2)^2\sigma^4})-\frac{\lambda^2
\ell(\ell+1)}{6(1-\lambda^2E^2)^2\sigma^4}  \Big ) \partial_{t}^2 +\frac{\ell(\ell+1)}{(1+\lambda
E)\sigma^2}\Big]\delta\phi^i_\ell =0. \eeq We
write $\de\phi^i_\ell =f^i_\ell e^{-iwt}\de x^i$ in the direction of $x^i$, then the equation
(51) becomes \beq \Big[ -\partial_{\sigma}^2 -\Big( \frac{1-\lambda
E}{1+\lambda E}(1+\lambda^2\frac{N^2 -1}{4(1-\lambda^2E^2)^2\sigma^4})-\frac{\lambda^2
\ell(\ell+1)}{6(1-\lambda^2E^2)^2\sigma^4}  \Big ) w^2 +\frac{\ell(\ell+1)}{(1+\lambda
E)\sigma^2}\Big]  f^i_\ell =0. \eeq To solve this equation we start, for simplicity, by considering small $\sigma$. The equation (52) is reduced to \beq \Big[
-\partial_{\sigma}^2 -\frac{\lambda^2w^2}{(1-\lambda^2E^2)^2\sigma^4}\Big(\frac{1-\lambda E}{1+\lambda E}\frac{N^2 -1}{4}-\frac{\ell(\ell+1)}{6} \Big )  \Big] f^i_\ell = \frac{1-\lambda E}{1+\lambda E}w^2f^i_\ell , \eeq 
with the potential $$V= \frac{-\lambda^2w^2}{(1-\lambda^2E^2)^2\sigma^4}\Big(\frac{1-\lambda E}{1+\lambda E}\frac{N^2 -1}{4}-\frac{\ell(\ell+1)}{6} \Big ).$$ The potential $V$ is quite zero for all $E$ and only at $\sigma\approx0.02$ that we see $V$ varies in terms of $E$ and goes up too fast to be close to zero as a constant function (Fig.4(a)).

The equation of motion (53) can be rewritten as follows \beq \Big[ -\frac{1+\lambda E}{1-\lambda
E}\partial_{\sigma}^2 -\Big( (1+\lambda^2\frac{N^2
-1}{4(1-\lambda^2E^2)^2\sigma^4})-\frac{1+\lambda E}{1-\lambda E}\frac{\lambda^2
\ell(\ell+1)}{6(1-\lambda^2E^2)^2\sigma^4}  \Big ) w^2 \Big] f^i_\ell = 0. \eeq We
change the coordinate to $\tilde{\sigma}=\sqrt{\frac{1-\lambda
E}{1+\lambda E}}w\sigma$ and the equation (54) is rewritten as \beq \Big[
\partial_{\tilde{\sigma}}^2+ 1+ \frac{\kappa^2}{\tilde{\sigma}^4}
\Big] f^i_\ell (\tilde{\sigma}) = 0, \eeq with
$$\kappa^2=w^4\lambda^2 \frac{3(1-\lambda E)^2 (N^2
-1)-2(1-\lambda^2 E^2)\ell(\ell+1)}{12(1+\lambda E)^2(1-\lambda^2E^2)^2}.$$ Then we
follow the suggestions of WKB by making a coordinate change; \beq
\beta(\tilde{\sigma})=\int\limits^{\tilde{\sigma}}_{\sqrt{\kappa}}dy\sqrt{1+
\frac{\kappa^2}{y^4}}, \eeq and \beq f^i_\ell
(\tilde{\sigma})=(1+\frac{\kappa^2}{\tilde{\sigma}^4})^{-\frac{1}{4}}\tilde{f}^i_\ell
(\tilde{\sigma}). \eeq Thus, the equation (55) becomes \beq\Big(
-\partial^{2}_{\beta}+V(\beta)\Big) \tilde{f^i}=0,\eeq with \beq
V(\beta)=\frac{5\kappa^2}{(\tilde{\sigma}^2
+\frac{\kappa^2}{\tilde{\sigma}^2})^3}. \eeq Then \beq
f^i_\ell=(1+\frac{\kappa^2}{\tilde{\sigma}^4})^{-\frac{1}{4}}e^{\pm
i\beta(\tilde{\sigma})}. \eeq Since we are dealing with small $\sigma$ case the obtained fluctuation becomes $$f^i_\ell=\frac{\tilde{\sigma}}{\sqrt{\kappa}}e^{\pm
i\beta(\tilde{\sigma})}.$$ This is the asymptotic wave function in the regions $\beta\rightarrow -\infty$, while around $\beta\sim 0$; i.e. $\tilde{\sigma}\sim\sqrt{\kappa}$, $f^i_\ell\sim 2^{-\frac{1}{4}}$. The variation of this fluctuation in terms of small $\sigma$ and the electric field is well shown in Fig.3(a) by considering the real part of the function. The variation of $f^i_\ell$ in terms of $\sigma$ has positive values and goes up as $\sigma$ goes up for all $E$ in general. The influence of $E$ on $f^i_\ell$ appears at $\sigma\approx0.2$.

Now, if $\sigma$ is too large the equation of motion (52) becomes
\beq \Big[ -\partial_{\sigma}^2 +\frac{\ell(\ell+1)}{(1+\lambda E)\sigma^2}\Big] f^i_\ell =  \frac{1-\lambda E}{1+\lambda E} w^2  f^i_\ell. \eeq

The fluctuation solution of this equation is \beq\bra{ll}f^i_\ell(\sigma)&=\alpha
\sqrt{\sigma}BesselJ\Big(\frac{1}{2}\sqrt{1+4\frac{\ell(\ell+1)}{1+\lambda E}},w\sigma\sqrt{\frac{1-\lambda E}{1+\lambda E}}\Big)\\&+\beta\sqrt{\sigma}BesselY\Big(\frac{1}{2}\sqrt{1+4\frac{\ell(\ell+1)}{1+\lambda E}},w\sigma\sqrt{\frac{1-\lambda E}{1+\lambda E}}\Big),\era\eeq with $\alpha$, $\beta$ are constants. The variation of this fluctuation in terms of large $\sigma$ and $E$ is given by Fig.3(b). The values of $f^i_\ell$ are negative and they are going down as $E$ going up.
%**************************
\begin{center}
\includegraphics[width=6in,height=4in]{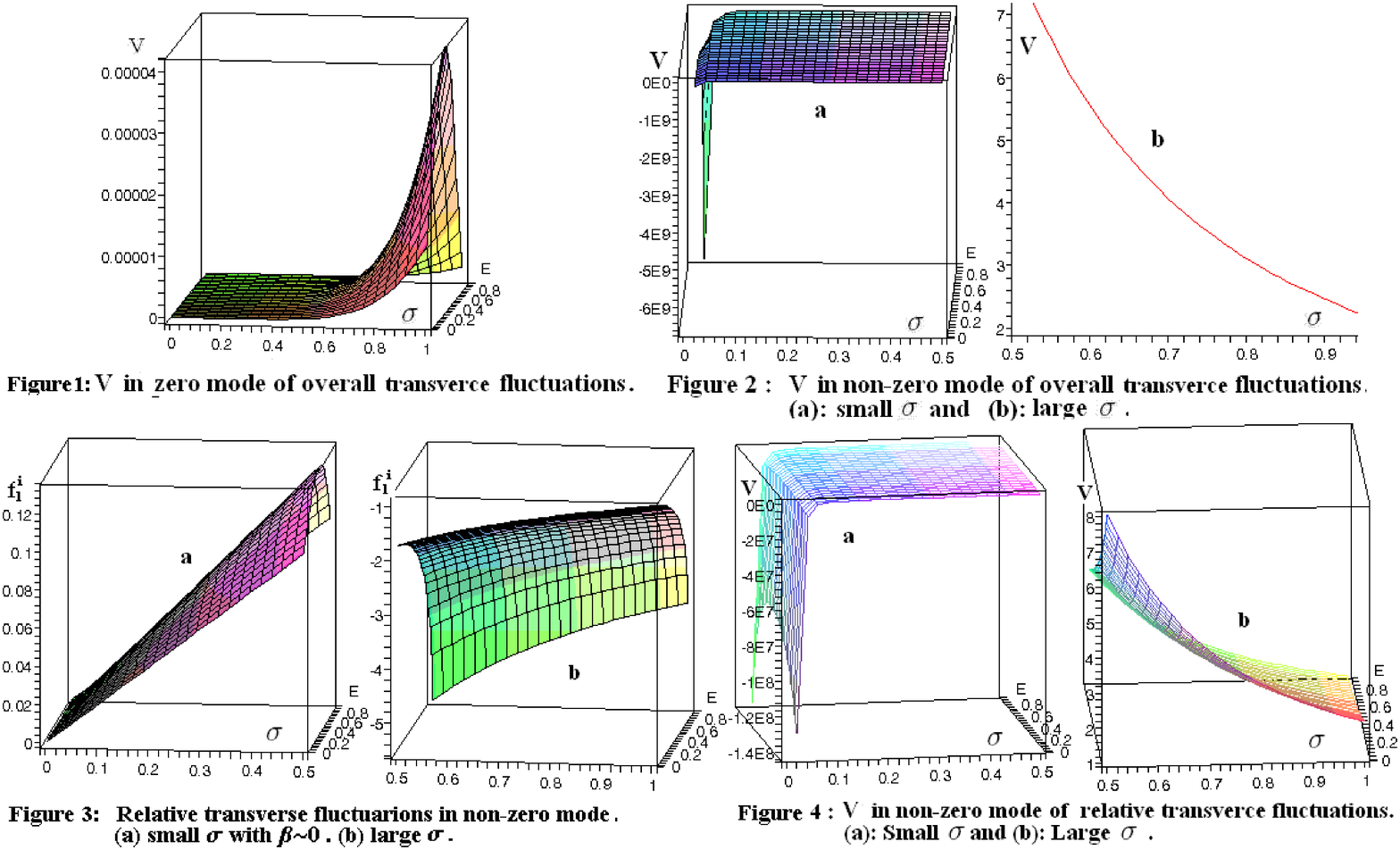}
\end{center}
%**************************
By dealing with the fluctuations (60) (Fig.3(a)) and (62) (Fig.3(b)) in small and large $\sigma$ regions respectively, we remark that is clear that we get different fluctuations from small to large $\sigma$ with a singularity at some stage of $\sigma$ and consequently the system is separated into two regions depending on the electric field.

The potential associated to (62) is $$V(\sigma)=\frac{\ell(\ell+1)}{(1+\lambda E)\sigma^2}.$$ Accordingly to Fig.4(b) describing the variation of $V$, we remark that, $V$ goes down as $\sigma$ goes up and more down as $E$ goes up; i.e. the potential becomes more small as the electric field appears between zero and $E\sim \frac{1}{\lambda}$. The potentials represented by the two figures Fig.4(a) and Fig.4(b) don't have an intersecting point at some stage of $\sigma$. This leads the system to get a singularity which supports the idea that the system is separated into two regions in non-zero modes of relative transverse fluctuations.

Consequently, the D1$\bot$D3 system has Neumann boundary conditions and this is more clear at the presence of electric field. This is proved through this section by discussing different modes and different directions of the fluctuations of the funnel solution and their associated potentials. In the figures representing these variations we set $w$, $\ell$ equals to the unit of $\lambda=1$ in all the treated equations and $N=10$.

\subsection{Overall Transverse Fluctuations in D1$\bot$D5 System}
\hspace{.3in}We extend this study to discuss the electrified fluctuations in D1$\bot$D5 system. we give the equations of motion of the fluctuations and their solutions. Then, we discuss the variation of the potential and the fluctuations in terms of electric field and the spatial coordinate.

We start by considering overall transverse fluctuations in zero mode and we let the non-zero modes with the relative transverse fluctuations to be discussed in the next project \cite{grav}. This type of fluctuations is given as $\delta \phi^m (\sigma,t)=f^m (\sigma,t)I_N$, $m=6,7,8$. Plugging this fluctuation into the full ($N,N_f$)-string action (15'), together with the funnel (16) the action is found to be \beq S=-NT_1 \int d^2\sigma \Big((1+\lambda E)A -\frac{1}{2}(1-\lambda^2 E^2 )\lambda^2 A(\dot{f}^m)^2+\frac{1}{2}(1+\lambda E)\lambda^2 (\partial_\sigma f^m)^2 +...\Big). \eeq where \beq A=(1+\frac{4R(\sigma)^4}{c\lambda^2 })^2 ,\eeq with the quadratic terms in $f^m$ were the only terms retained in the action. The linearized equation of motion of the fluctuation is then \beq\Big[ (1-\lambda E)\Big( 1+\frac{4R(\sigma)^4}{c\lambda^2 }\Big)^2 \partial^{2}_{t}-\partial^{2}_{\sigma}\Big] f^m =0.\eeq We consider small $R$ which is given by (17) in the second section. We insert its expression in the last equation and the equation of motion becomes \beq\Big[ (1-\lambda E)\Big( 1+\frac{n^2 \lambda^2 }{16(1-\lambda^2E^2)^2\sigma^4}\Big)^2 \partial^{2}_{t}-\partial^{2}_{\sigma}\Big] f^m =0\eeq for large
$n$.

Let's consider the fluctuation in the following form \beq
f^m =\phi(\sigma)e^{-iwt}\delta x^m, \eeq with $\delta x^m$, $m=6,7,8$, the
direction of the fluctuation. The equation (66) becomes
\beq\Big[-\partial^{2}_{\sigma} -w^2 (1-\lambda E)(\frac{n^2
\lambda^2 }{8(1-\lambda^2E^2)^2\sigma^4}+\frac{n^4 \lambda^4 }{16^2(1-\lambda^2E^2)^4\sigma^8}) \Big]
\phi =w^2 (1-\lambda E)\phi.\eeq The potential  of this system is \beq V= -w^2 (1-\lambda E)(\frac{n^2
\lambda^2 }{8(1-\lambda^2E^2)^2\sigma^4}+\frac{n^4 \lambda^4 }{16^2(1-\lambda^2E^2)^4\sigma^8})\eeq depending on the electric field $E$ with $E\in [0,\frac{1}{\lambda}[$.
\subsubsection{Small $\sigma$ Limit}
The equation (68) is complicated and to simplify the calculations we start by considering the small $\sigma$ and $\frac{1}{\sigma^8}$ dominates in (68) and (69). We then discuss the equation
\beq\Big[-\partial^{2}_{\sigma} -w^2 (1-\lambda E)\frac{n^4 \lambda^4 }{16^2(1-\lambda^2E^2)^4\sigma^8}\Big]
\phi =w^2 (1-\lambda E)\phi,\eeq and the potential is reduced to \beq V= -w^2 (1-\lambda E)\frac{n^4 \lambda^4}{16^2 (1-\lambda^2E^2)^4\sigma^8}.\eeq

As shown in Fig.5(a), the potential $V$ tends to $-\infty$ until some values of $E$ when $E$ and $\sigma$ are close to zero, and once $E$ is close to the inverse of $\lambda$ the potential is zero for all small $\sigma$. We consider in this case $\sigma\in]0,0.5]$ in the unit of $\lambda$ with $\lambda=1$, $w=1$, $n=10$ and $E\in[0,1[$.

To solve the differential equation (70), we consider the total differential on the fluctuation. Let's denote $\partial_{\sigma}\phi\equiv \phi'$. Since $\phi$ depends only on $\sigma$ we find $\frac{d\phi}{d\sigma}=\partial_{\sigma}\phi$. We rewrite the equation (70) in this form \beq \frac{1}{\phi}\frac{d\phi'}{d\sigma}=-w^2 (1-\lambda E)[\frac{n^4 \lambda^4 }{16^2(1-\lambda^2E^2)^4\sigma^8}+1].\eeq An integral formula can be written as follows \beq \int\limits_{0}^{\phi'}\frac{d\phi'}{\phi}=-\int\limits_{0}^{\sigma} w^2 (1-\lambda E)[\frac{n^4 \lambda^4 }{16^2(1-\lambda^2E^2)^4\sigma^8}+1]d\sigma,\eeq which gives \beq \frac{\phi'}{\phi}=-w^2 (1-\lambda E)[-\frac{n^4 \lambda^4 }{16^2(1-\lambda^2E^2)^4\times 7\sigma^7}+\sigma]+\alpha.\eeq We integrate again the following \beq\int\limits_{0}^{\phi}\frac{d\phi}{\phi}=-\int\limits_{0}^{\sigma}(w^2 (1-\lambda E)[-\frac{n^4 \lambda^4 }{16^2\times 7(1-\lambda^2E^2)^4\sigma^7}+\sigma]+\alpha) d\sigma.\eeq We get \beq\ln\phi=-w^2 (1-\lambda E)[-\frac{n^4 \lambda^4 }{16^2\times 42(1-\lambda^2E^2)^4\sigma^6}+\frac{2}{\sigma^2}]+\alpha\sigma+\beta,\eeq and the fluctuation in small $\sigma$ region is found to be \beq \phi (\sigma)=\beta e^{-w^2 (1-\lambda E)[-\frac{n^4 \lambda^4 }{16^2\times 42(1-\lambda^2E^2)^4\sigma^6}+\frac{\sigma^2}{2}]+\alpha\sigma},\eeq with $\beta$ and $\alpha$ are constants.

We plot the progress of the obtained fluctuation in (Fig.6(a)). First we consider the constants $\beta=1=\alpha$, then the small spatial coordinate in the interval $[0,0.5]$ with the unit of $\lambda=1$, $w=1$ and $n=4$. As above the electric field is in $[0,1[$. We see that at the absence of the electric field there is no fluctuations at all and this phenomenon continues for the small values of $E$. When $E\approx0.5$ the fluctuation appears from $\sigma=0.15$ and goes down as $\sigma$ and $E$ go up.

\subsubsection{Large $\sigma$ Limit}
In the large $\sigma$ case, the equation (68) becomes \beq\Big[-\partial^{2}_{\sigma} -w^2 (1-\lambda E)\frac{n^2\lambda^2 }{8(1-\lambda^2E^2)^2\sigma^4}\Big]\phi =w^2 (1-\lambda E)\phi\eeq and the potential is \beq V= -w^2 (1-\lambda E)\frac{n^2\lambda^2 }{8(1-\lambda^2E^2)^2\sigma^4}.\eeq By plotting the progress of this potential (Fig.5(b)) we consider the large spatial coordinate in the interval $[0.5,1]$ and $E\in[0,1[$ in the unit of $\lambda=1$, $w=1$ with $n=10$. The obtained figure shows that $V$ has in general higher values than the ones obtained in small $\sigma$ case (Fig.5(a) describing (71)). Specially, for the first values of $\sigma$, $V$ goes up from negative values to be close to zero for almost values of $E$ until $E$ is close to $\frac{1}{\lambda}$, approximately from $E=0.8$ where $V\approx -0.02$, we remark that $V$ has small variation in [0.8,1] region of $\sigma$. By contrary, in figure 5(a), when $\sigma=0.5$ which is the last value of $\sigma$ in that case we find $V$ is already zero for all $E$. Consequently, these two potentials (71) and (79) show a big gap to go from one system to other that they describe, meaning that our system is separated into two regions; small and large $\sigma$ depending on $E$.

Now, we should solve the equation of motion of the relative transverse fluctuations (78), in the case of large $\sigma$. We start by defining a new coordinate
$$\tilde{\sigma}^2= w^2 (1-\lambda E)\sigma^2$$ and (78) becomes \beq\Big( 1+\frac{n^2 \lambda^2 w^4 (1-\lambda E)^2}{8(1-\lambda^2 E^2)^2\tilde{\sigma}^4} +\partial^{2}_{\tilde{\sigma}}\Big) \phi(\tilde{\sigma}) =0,\eeq with the potential is \beq
V(\tilde{\sigma})=\frac{\kappa^2}{\tilde{\sigma}^4}, \eeq and
$$\kappa^2=\frac{n^2 \lambda^2 w^4 (1-\lambda E)^2}{8(1-\lambda^2 E^2)^2}.$$ The equation
(80) is a Schr\"odinger equation for an attractive singular
potential $\propto\tilde{\sigma}^{-4}$ and depends on the single
coupling parameter $\kappa$ with constant positive Schr\"odinger
energy. The solution is then known by making the following
coordinate change \beq
\chi(\tilde{\sigma})=\int\limits^{\tilde{\sigma}}_{\sqrt{\kappa}}
dy\sqrt{1+\frac{\kappa^2}{y^4}}, \eeq and \beq
\Phi=(1+\frac{\kappa^2}{\tilde{\sigma}^4})^{-\frac{1}{4}}\tilde{\Phi}.
\eeq Thus, the equation (80) becomes \beq\Big(
-\partial^{2}_{\chi}+V(\chi)\Big) \tilde{\Phi}=0,\eeq with \beq
V(\chi)=\frac{5\kappa^2}{(\tilde{\sigma}^2
+\frac{\kappa^2}{\tilde{\sigma}^2})^3}.\eeq Then, the fluctuation is
found to be \beq
\Phi=(1+\frac{\kappa^2}{\tilde{\sigma}^4})^{-\frac{1}{4}}e^{\pm
i\chi(\tilde{\sigma})}. \eeq This fluctuation has the following
limit; since we are in large $\sigma$ region $\Phi\sim e^{\pm i\chi(\tilde{\sigma})}$. This is the asymptotic wave function in the regions $\chi\rightarrow +\infty$, while around $\chi\sim 0$;
i.e. $\tilde{\sigma}\sim\sqrt{\kappa}$, $\Phi\sim 2^{-\frac{1}{4}}$. Owing to the plotting of the progress of this fluctuation given by Fig.6(b), by considering the real part of the function, we remark that $\Phi$ goes down fast as $E$ goes up for all $\sigma$. When $\sigma=0.5$ the fluctuation gets different values for all $E$ compared to the values gotten in small $\sigma$ region by (77) (Fig.6(a)). These tow figures show that the fluctuations of fuzzy funnel of D1$\bot$D5 branes have a singularity at some stage of $\sigma$ separating the system into two regions; small and large $\sigma$.
%**************************
\begin{center}
\includegraphics[width=4in,height=4in]{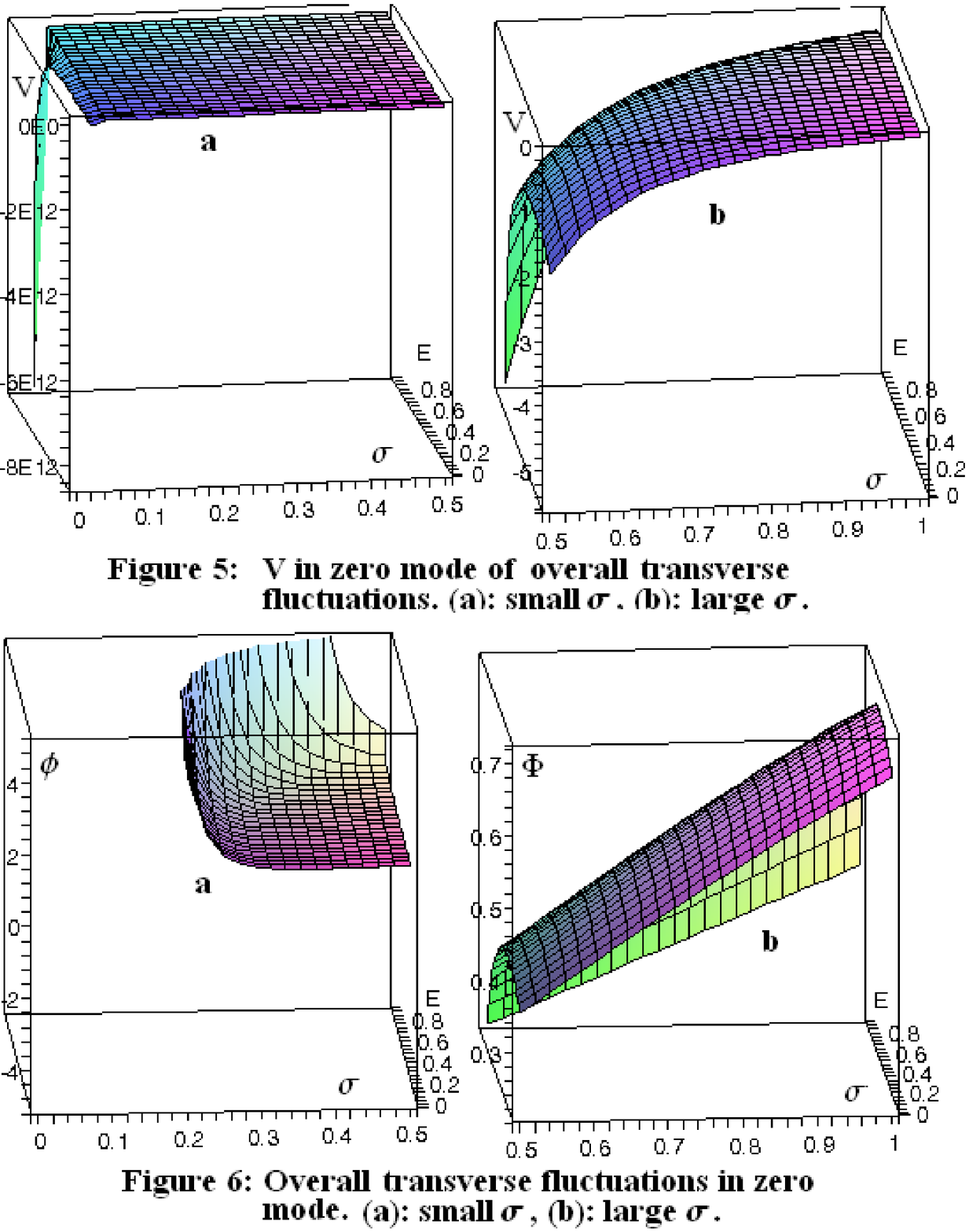}
\end{center}
%**************************
\section{Discussion and Conclusion}
\hspace{.3in}In this work, we showed that certain excitations of D1$\bot$D3 and D1$\bot$D5 systems can be shown to obey Neumann boundary conditions. We considered the non-abelian BI dynamics of the dyonic string such that the electric field $E$ has a limited value. The limit of $E$ attains a maximum value \beq E_{max}=T_1=\frac{1}{\lambda}\eeq (for simplicity we dropped $2\pi$ in all the calculations). This limiting value arises because if $E>\frac{1}{\lambda}$ the action ceases to make physical sense \cite{nphys}. The system becomes unstable. Since The string effectively carries electric charges of equal sign at each of its endpoints, as E increases the charges start to repel each other and stretch the string. For E larger than the critical value (87), the string tension $T_1$ can no longer hold the strings together. 

In this context, by considering $E\in[1,\frac{1}{\lambda}[$ in D1$\bot$D3 and D1$\bot$D5 branes, we treated the fluctuations of the fuzzy funnel solutions and discussed the associated potentials $V$ in terms of the electric field $E$ and the spatial coordinate $\sigma$. We considered the unit of $\lambda$ in all the figures representing the variations. We limited $\sigma$ to be in the interval [0,0.5] for small $\sigma$ and [0.5,1] for large $\sigma$.

Concerning D1$\bot$D3 system, we gave the variation of $V$ in zero mode of overall transverse fluctuations in Fig.1; this figure shows that the system is looking like separated into two regions depending on the electric field. The potential is stable for $E$ varies from 0 until 0.7 and then $V$ goes up quickly as $E$ close to $\frac{1}{\lambda}$. Then we dealt with the general case of the overall transverse fluctuations which is the non-zero modes. In this case, the idea that the system is divided into two regions appears more clear. We gave the figure representing the potential in Fig.2; we see that at $\sigma=0.5$ the potential gets a singularity. We continued to treat the other kind of the fluctuations, it's the relative transverse fluctuations. These fluctuations are represented  by Fig.3 and the associated potentials by Fig.4; we obtain the same remark that both fluctuations and potential have a singularity at $\sigma=0.5$. This supported the idea that the system is separated into two regions.  

We extended our study to the case of higher dimensions. We treated the electrified fluctuations of D1$\bot$D5 branes and we studied only the zero mode of overall transverse fluctuations. All other cases will be discussed in the coming work. We notice that when the electric field is going up and down the potential of the system is changing and the appearance of the singularity is more clear (Fig.5) and we have the same remarks for the fluctuations of fuzzy funnel solutions as well (Fig.6) which cause the division of the system into tow regions depending on small and large $\sigma$ and also on $E$.

Consequently, the end point of the dyonic strings moves on the brane which means we have Neumann boundary conditions in D1$\bot$D3 and D1$\bot$D5 branes. The physical interpretation is that a string attached to the D3 and D5 branes manifests itself as an electric charge, and the waves on the string cause the end point of the string to freely oscillate. Thus, we realize Polchinski's open string Neumann boundary conditions dynamically by considering non-abelian BI action in D1$\bot$D3 and D1$\bot$D5 systems.

In the coming work \cite{grav}, we will focus on other interesting investigations concerning the perturbations propagating on a dyonic string in the supergravity background \cite{supergravComp,dual} of an orthogonal p-brane. We will discuss the relative transverse fluctuations in dyoinc D1-D5 system in flat background and supergravity background.


\begin{thebibliography}{99}

\bibitem{fluc} J. Douari and A. H. Ali, Prog. in Phys., 3 (2007) 26-32; hep-th/0610288.

\bibitem{cm} N. R. Constable, R.C. Myers, O. Tafjord, JHEP 0106 (2001) 023, hep-th/0102080; R. C. Myers, JHEP 9912 (1999) 022, hep-th/9910053; J. Kluson, JHEP 0011 (2000) 016, hep-th/0009189.

\bibitem{cm2} Y. Chen, B. Hou, B.-Y. Hou, Nucl. Phys. B 638 (2002) 220, hep-th/0203095;
\bibitem{cm3} J. Douari, Phys. Lett. A367 (2007) 52-56.

\bibitem{BI} J. Polchinski, Tasi Lectures on D-branes,
hep-th/9611050; R. Leigh, Mod. Phys. Lett. A4 (1989) 2767.
\bibitem{InterBran1} C. G. Callan and J. M. Maldacena, Nucl. Phys.
B513 (1998) 198, hep-th/9708147.
\bibitem{InterBran2} G. W. Gibbons, Nucl. Phys. B514 (1998) 603, hep-th/9709027; P. S. Howe, N. D. Lambert and P. C. West, Nucl. Phys. B515(1998) 203, hep-th/9709014; T. Banks,
W. Fischler, S. H. Shenker and L. Susskind, Phys. Rev. D55 (1997)
5112, hep-th/9610043; D. Kabat and W. Taylor, Adv. Theor. Math. Phys.
2 (1998) 181, hep-th/9711078; S. Rey, hep-th/9711081; N. R. Constable, R. C. Myers and O. Tafjord,  hep-th/0105035; R. Blumenhagen, M. Cvetic, P. Langacker, and G. Shiu, hep-th/0502005.
\bibitem{dd3} J. Douari, Phys. Lett. B656 (2007) 233-242; hep-th/0610156.
\bibitem{fun} D. Brecher, Phys. Lett. B 442 (1998)
117, hep-th/9804180; P. Cook, R. de Mello Koch and J. Murugan, Phys.
Rev. D 68 (2003) 126007, hep-th/0306250; N. R. Constable, R. C.
Myers and O. Tafjord, Phys. Rev. D 61 (2000) 106009, hep-th/9911136.
\bibitem{dual} N. R. Constable, R. C. Myers and O. Tafjord, Phys. Rev. D61, 106009 (2000),
hep-th/9911136.
\bibitem{NBCBI} K. G. Savvidy and G. K. Savvidy, Nucl. Phys. B561(1999) 117–124; hep-th/9902023.
\bibitem{fuzfun} N. R. Constable,
R. C. Myers and O. Tafjord, Phys. Rev. D61, 106009 (2000),
hep-th/9911136; R. Bhattacharyya and R. de Mello Koch,
hep-th/0508131; C. Papageorgakis , S. Ramgoolam , N. Toumbas,
hep-th/0510144.
\bibitem{fluct} S. Lee, A. Peet and L. Thorlacius, Nucl. Phys. B514
(1998) 161, hep-th/9710097; D. Kastor and J. Traschen, Phys. Rev.
D61 (2000) 024034, hep-th/9906237.
\bibitem{d1d3}J. Douari, Phys. Lett. B644 (2007) 83-87; hep-th/0603037.
\bibitem{NBI}O. Obregon, hep-th/0505121.
\bibitem{9911136} N. R. Constable, R. C. Myers, O. Tafjord,
Phys. Rev. D61 (2000) 106009, hep-th/9911136.
\bibitem{Gib} G. W. Gibbons, Nucl. Phys. B514 (1998) 603-639, hep-th/9709027.
\bibitem{STr} A. A. Tseytlin, Nucl. Phys. B501 (1997) 41, hep-th/9701125; A. A. Tseytlin, Imperial/TP/98-99/67, hep-th/9908105; R. C. Myers, JHEP 9912 (1999) 022, hep-th/9910053. 
\bibitem{supergravComp} S. Lee, A. Peet and L. Thorlacius, Nucl.
Phys. B514, 161 (1998), hep-th/9710097; D. Kastor and J. Traschen,
Phys. Rev. D61, 024034 (2000), hep-th/9906237; N. R. Constable, R.C.
Myers and O. Tafjord, Phys. Rev. D61, 106009 (2000), hep-th/9911136.
\bibitem{funnelSoluD5} J. Castelino, S. Lee and W. Taylor, Nucl.
Phys. B526, 334 (1998), hep-th/9712105; H. Grosse, C. Klimcik and P.
Presnajder, Commun. Math. Phys. 180, 429 (1996), hep-th/9602115.
\bibitem{nphys} E. S. Fradkin and A. A. Tseytlin, Phys. Lett. B163 (1985) 295; R. J. Szabo, "An Introduction to String Theory and D-brane Dynamics", Imperial College Press, London, 2004.
\bibitem{grav} J. Douari, work in progress

\end{thebibliography}
\end{document}